\documentclass{article}

\usepackage{arxiv}

\usepackage[utf8]{inputenc} 
\usepackage[T1]{fontenc}    
\usepackage{hyperref}       
\usepackage{url}            
\usepackage{booktabs}       
\usepackage{amsfonts}       
\usepackage{nicefrac}       
\usepackage{microtype}      
\usepackage{graphicx}
\usepackage[numbers]{natbib}
\usepackage{doi}
\usepackage{amsmath, amssymb}
\usepackage{algorithm}
\usepackage{algpseudocode}
\usepackage{tikz}
\usepackage{float}
\usepackage{cleveref}
\usepackage{multirow}
\usepackage{array}
\usepackage{siunitx}
\usepackage{makecell}
\usepackage{placeins}

\usetikzlibrary{positioning, shapes.geometric, arrows.meta, fit, calc, decorations.pathreplacing}
\tikzstyle{box} = [rectangle, rounded corners, minimum width=3.5cm, minimum height=2.5cm, text centered, draw=black, fill=gray!10]
\tikzstyle{NNnode}=[thick,draw=black,fill=black!20,circle,minimum size=6]
\tikzstyle{arrow} = [very thick, ->, >=Latex, line width=1.5pt]
\tikzstyle{oval} = [ellipse, minimum width=7cm, minimum height=7cm, text centered, draw=black, fill=gray!10]
\tikzstyle{smalloval} = [ellipse, minimum width=5.5cm, minimum height=4.5cm, text centered, draw=black, fill=gray!10]
\tikzstyle{smallplot} = [draw=none]

\newcommand{\tikzmark}[1]{%
    \tikz[remember picture, overlay] \coordinate (#1);%
}

\title{Publishing Neural Networks in Drug Discovery Might Compromise Training Data Privacy}

\date{October 18, 2024}	

\author{ \href{https://orcid.org/0009-0005-6420-2175}{\includegraphics[scale=0.06]{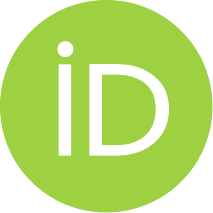}\hspace{1mm}Fabian P. Krüger$^{1,2,3}$} \\
        \And \href{https://orcid.org/0000-0003-4138-0508}{\includegraphics[scale=0.06]{orcid.pdf}\hspace{1mm}Johan Östman$^{4}$} \And 
        \href{https://orcid.org/0000-0002-7271-0824}{\includegraphics[scale=0.06]{orcid.pdf}\hspace{1mm}Lewis Mervin$^{5}$} \And 
        \href{https://orcid.org/0000-0002-6855-0012}{\includegraphics[scale=0.06]{orcid.pdf}\hspace{1mm}Igor V. Tetko$^{3}$} \And 
        \href{https://orcid.org/0000-0003-4970-6461}{\includegraphics[scale=0.06]{orcid.pdf}\hspace{1mm}Ola Engkvist$^{1,6}$}
        \AND
        $^1$~AstraZeneca R\&D\\ 
        Discovery Sciences \\
        Molecular AI \\
        431 83 Mölndal, Sweden  \\
        \And
        $^2$~Technical University of Munich \\ TUM School of Computation, \\ Information and Technology \\ 
        Department of Mathematics \\
        80333 Munich, Germany 
        \And
        $^3$~Helmholtz Munich \\
        Molecular Targets and Therapeutics Center \\
        Institute of Structural Biology \\
        85764 Neuherberg, Germany 
        \And
        $^4$~AI Sweden \\
        41756 Gothenburg, Sweden
        \And
        $^5$~AstraZeneca R\&D\\ 
        Discovery Sciences \\
        Molecular AI \\
        CB2 0AA Cambridge, UK  \\
        \And
        $^6$~Chalmers University of Technology \\
        Department of Computer Science \\ and Engineering \\
        412 96 Gothenburg, Sweden
    }

\renewcommand{\shorttitle}{\textit{arXiv}}

\hypersetup{
pdftitle={PUBLISHING NEURAL NETWORKS IN DRUG DISCOVERY
COMPROMISES TRAINING DATA PRIVACY},
pdfsubject={arxiv pre-print},
pdfauthor={Fabian Krüger},
pdfkeywords={membership inference attack, privacy, drug discovery, cheminformatics, QSAR, machine learning},
}

\begin{document}
\maketitle

\begin{abstract}
This study investigates the risks of exposing confidential chemical structures when machine learning models trained on these structures are made publicly available. We use membership inference attacks, a common method to assess privacy that is largely unexplored in the context of drug discovery, to examine neural networks for molecular property prediction in a black-box setting. Our results reveal significant privacy risks across all evaluated datasets and neural network architectures. Combining multiple attacks increases these risks. Molecules from minority classes, often the most valuable in drug discovery, are particularly vulnerable. We also found that representing molecules as graphs and using message-passing neural networks may mitigate these risks. We provide a framework to assess privacy risks of classification models and molecular representations. Our findings highlight the need for careful consideration when sharing neural networks trained on proprietary chemical structures, informing organisations and researchers about the trade-offs between data confidentiality and model openness.
\end{abstract}


\section*{Main}

The use of neural networks has gained significant traction in early drug discovery, with organisations increasingly relying on these models for a range of important modelling tasks \cite{chen2018rise}. One of the most common applications is the prediction of molecular properties \cite{muratov2020qsar, dara2022machine}. The performance of these models is heavily dependent on the quality and quantity of available datasets \cite{muratov2020qsar}. However, generating these datasets in drug discovery is an expensive and resource-intensive process, often requiring significant investment in both time and money \cite{vamathevan2019applications}. As a result, organisations are highly protective of their data, as they have invested significant resources in generating the proprietary datasets  and are accordingly reluctant to make this information publicly available.

While organisations are interested in keeping their proprietary datasets private due to the significant investments involved, they still recognise the value of engaging with the broader drug discovery community and artificial intelligence (AI) communities \cite{oldenhof2023industry}. In the AI research field, it is common practice to share models through open-source platforms or alternatively to offer them as secure web services, fostering collaboration and innovation \cite{zuckerberg2024opensource}. This interaction is mutually beneficial, as it allows for the refinement and validation of models while also advancing the field as a whole \cite{shrestha2023building}. However, this type of collaboration inevitably raises concerns about data security, an issue of growing importance in AI research \cite{murdoch2021privacy}. As organisations seek to balance the advantages of community engagement with the need to protect valuable data, the issue of privacy is becoming increasingly important.

In this work, we adopt an interdisciplinary approach that bridges the fields of drug discovery and data privacy research. This bridge has largely been missing and we firmly believe that there are great opportunities for scientific progress by bringing the two fields closer to each other. To empirically evaluate the privacy of machine learning models, membership inference attacks have become the most widely used method \cite{shokri2017membership, murakonda2020ml, carlini2022membership}. These attacks can be conceptualized as a privacy game, where the adversary seeks to determine whether a specific sample was part of the model's training data (\Cref{alg:MIA}). There are various levels of information the adversary might have access to regarding the model \cite{salem2023sok}. In our study, we focus on the so-called black-box scenario, where the adversary is provided with the output logits of the trained model, rather than the model's weights, which would correspond to a white-box scenario. This black-box scenario is similar to making machine learning models available as web services. 

\begin{algorithm}
\caption{\textbf{Membership Inference Attack.} This algorithm formalizes the membership inference attack game we use to evaluate the privacy of our neural networks. The attack assumes knowledge about the underlying data distribution (chemical space) $\Pi$ from which the training dataset is sampled.  Given an adversary $A$, a training algorithm $T$, and the data distribution $\Pi$, the process involves sampling points from the data distribution, training a model on these samples, and then using the adversary to infer whether a specific data point (chemical structure) was part of the training set or not. The algorithm tests the adversary's ability to distinguish between data points sampled from the training set and those not included, thereby evaluating potential information leakage from the model.}

\label{alg:MIA}
\begin{algorithmic}[1]
    \State \textbf{Input:} Adversary $A$, Training Algorithm $T$, Data distribution $\Pi$
    \State Sample $n$ points from $\Pi$: $D \sim \Pi^n$
    \State Train model using $T$ on $D$: $f_{\theta} \leftarrow T(D)$
    \State Flip a coin: $b \sim \{0, 1\}$
    \If{$b = 0$}
        \State Sample $z \sim D$
    \Else
        \State Sample $z \sim \Pi(\cdot \mid z \notin D)$
    \EndIf
    \State Let $A$ guess $b$: $\tilde{b} \leftarrow A(T, \Pi, z, f_{\theta}(z))$
\end{algorithmic}
\end{algorithm}

Building on the growing body of research on membership inference attacks, Hu et al. conducted an extensive survey, highlighting that they have been studied in the domains of image data, text data, tabular data, as well as node classification in graph data \cite{hu2022membership}. Among the different implementations of attacks, likelihood ratio attacks (LiRA) and robust membership inference attacks (RMIA) have been shown to be the most effective in identifying training data samples, setting state-of-the-art performance benchmarks for the most commonly used benchmark datasets \cite{carlini2022membership, zarifzadeh2024low}. Despite the growing interest in membership inference attacks, their application to molecular property prediction in drug discovery remains largely unexplored. To the best of our knowledge, Pejo et al. conducted the only study about membership inference attacks in the context of molecular property prediction, but they focused on federated learning scenarios using attacks tailored to this approach \cite{pejo2022collaborative}. The broader implications and potential risks of membership inference attacks in molecular property prediction, particularly in traditional centralised machine learning models, still require investigation.

\begin{figure}[h!]
    \centering
    \includegraphics[width=\textwidth]{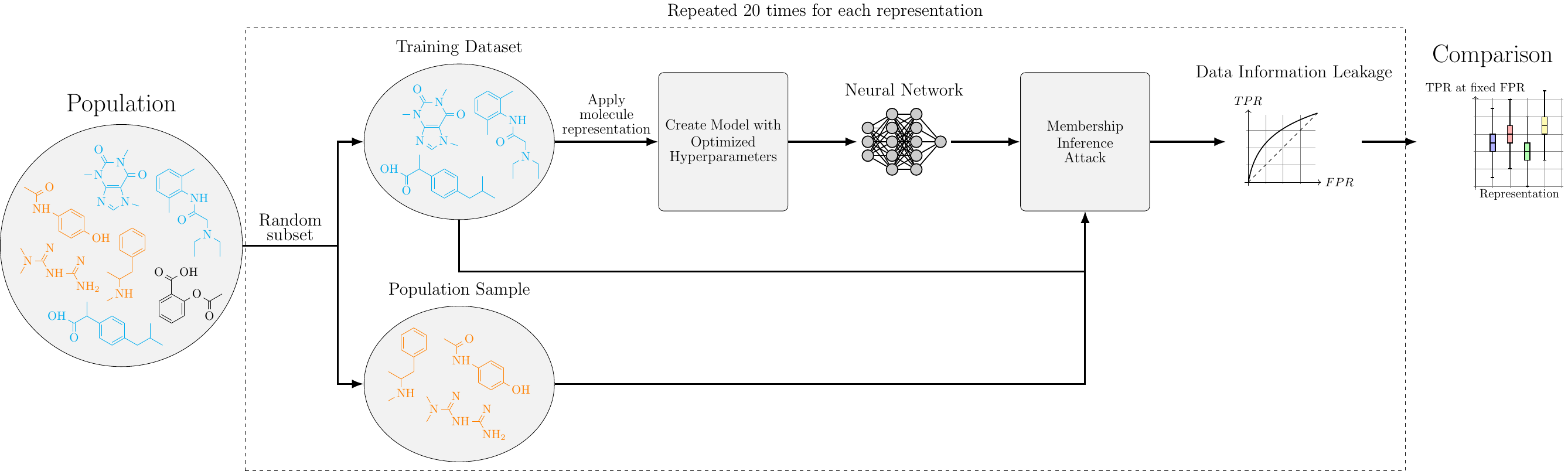} 
    \caption{\textbf{Overview of our workflow to evaluate privacy risks of neural network for molecular property prediction.} Two random, non-overlapping subsets are created from each dataset. One subset is transformed into the desired molecular representation and used to train a neural network, optimised through Bayesian hyperparameter tuning \cite{bergstra2011algorithms}. We then apply membership inference attacks (\Cref{alg:MIA}) to determine if chemical structures in the training data can be distinguished from those in the other subset. We evaluate this using two different attack implementations. This process is repeated 20 times for each dataset and molecular representation. We assess the results by analyzing true positive rates at fixed false positive rates, comparing them to random guessing, and examining the impact of the  molecular representations.}
    \label{fig:workflow}
\end{figure}

In this study, we provide the first comprehensive analysis of membership inference attacks against neural networks trained to predict molecular properties. We thereby highlight the risk that releasing machine learning models may expose proprietary chemical structures to the public, a challenge that organisations, for instance, must consider. To our knowledge this is the first study to investigate how different molecular representations affect the privacy of the resulting models. Additionally, we create a framework where the privacy risks of classification model architectures and representation algorithms can be assessed and compared. A scheme of our workflow is described in \Cref{fig:workflow}. Our study also explores whether different membership inference attacks can be used together, and we present some characteristics of the identified chemical structures that provide insights into the specific privacy risks. The approaches and findings of this study have relevance beyond the pharmaceutical sector, offering applicability to any field that relies on predictions of molecular properties, such as materials science or toxicology. Our framework also allows for the systematic assessment of privacy threats associated with predictive models in these fields.

\section*{Results}

In this section, we present the results of membership inference attacks on different neural networks trained on different datasets for specific tasks: Blood-Brain Barrier crossing (BBB) to predict the ability of molecules to cross the blood-brain barrier \cite{martins2012bayesian}, Ames mutagenicity prediction (Ames) to assess potential mutagenicity \cite{hansen2009benchmark, xu2012silico}, DNA Encoded Library enrichment (DEL) to analyse enrichment \cite{lim2022machine}, and inhibition of the potassium ion channel encoded by the human ether-à-go-go-related gene (hERG) to assess cardiac toxicity risks \cite{du2011hergcentral}. The datasets differ in size with BBB and Ames being relatively small (859 and 3,264 training data molecules) and DEL and hERG being relatively large (48,837 and 137,853 training data molecules).  We explore the potential of combining different attacks to identify additional molecules contained in the training data. We also investigate whether the identified molecules have distinct properties that distinguish them from the rest of the training data. Finally, we provide a detailed example of a specific attack to illustrate our findings.

\subsection*{Membership inference attacks}

\begin{figure}[h!]
    \centering
    \includegraphics[width=\textwidth]{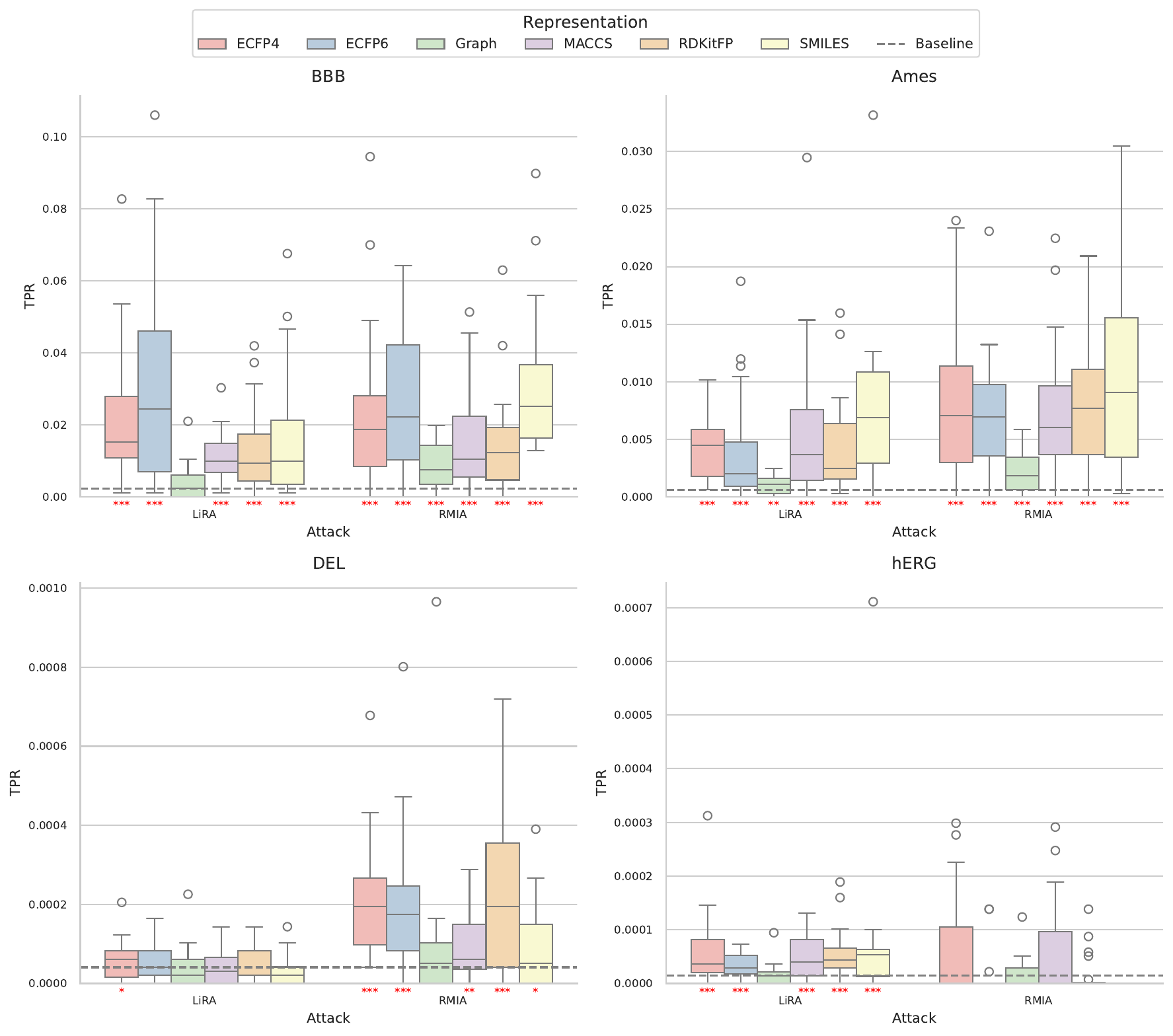} 
    \caption{\textbf{True positive rates for identifying training data molecules at a false positive rate of 0.} The distributions of 20 experimental repetitions are shown for each representation and dataset, for both the likelihood ratio attack (LiRA) and the robust membership inference attack (RMIA). Distributions with significantly higher true positive rates than the baseline are indicated by red stars. A single star represents a p-value less than 0.05, two stars represent a p-value less than 0.01, and three stars represent a p-value less than 0.001. Training dataset sizes (total amount of positives) are: 859 molecules for the blood-brain barrier permeability dataset; 3,264 for the Ames mutagenicity prediction dataset; 48,837 for the DNA-encoded library enrichment dataset; and 137,853 for the hERG channel inhibition dataset.}
    \label{fig:FPR0}
\end{figure}

We wanted to see if we could identify whether a molecule was part of the training data from querying a neural network and analysing its outputs. To achieve this, we used two different membership inference attacks: likelihood ratio attacks (LiRA) and robust membership inference attacks (RMIA) \cite{carlini2022membership, zarifzadeh2024low}. We evaluated their ability to distinguish between molecules in the training data and those outside it by measuring the true positive rate (TPR) at a false positive rate (FPR) of 0. In this context, we refer to molecules that were part of the training data as positives. Evaluating membership inference attacks at low FPRs was recommended by Carlini et al. \cite{carlini2022membership}. Here we examine the TPR at an FPR of 0, which is the most conservative approach. For models trained on smaller datasets, we observed significantly higher TPRs than would be observed when randomly guessing if the chemical structure was part of the training dataset (\Cref{fig:FPR0}). For example, in the blood-brain barrier crossing dataset, median TPRs were between 0.01 and 0.03 for most representations, corresponding to the identification of between 9 and 26 of the 859 training molecules. The baseline in our experimental setup for identifying molecules by chance is identifying 2 molecules of the training data (See Supplementary Information for a comprehensive derivation of this baseline). Models trained on larger datasets also showed significantly high TPRs, but only for one of the attacks, which varied between datasets (\Cref{fig:FPR0}). The observed TPRs decreased with increasing dataset size.

To verify the consistency of our trends, we repeated our analysis of the TPR at an FPR of $10^{-3}$, as shown in Supplementary Figure 1. We observed similar trends at this FPR. One notable difference was that RMIA always performed at least as well as LiRA across every dataset and representation. Specifically, RMIA was significantly better in half of the cases. For the other half, no significant difference was observed. In addition, even for the larger datasets, RMIA consistently provided higher TPRs than the baseline. We also investigated the corresponding ROC curves for all datasets and representations, which show our trends are consistent even for larger FPRs (Supplementary Figure 2). The high TPRs across all four datasets at both FPRs indicate significant information leakage, showing that chemical structures from the training data can be identified. The amount of information leakage seems to be higher for models trained on smaller datasets.

\begin{figure}[h!]
    \centering
    \includegraphics[width=0.7\textwidth]{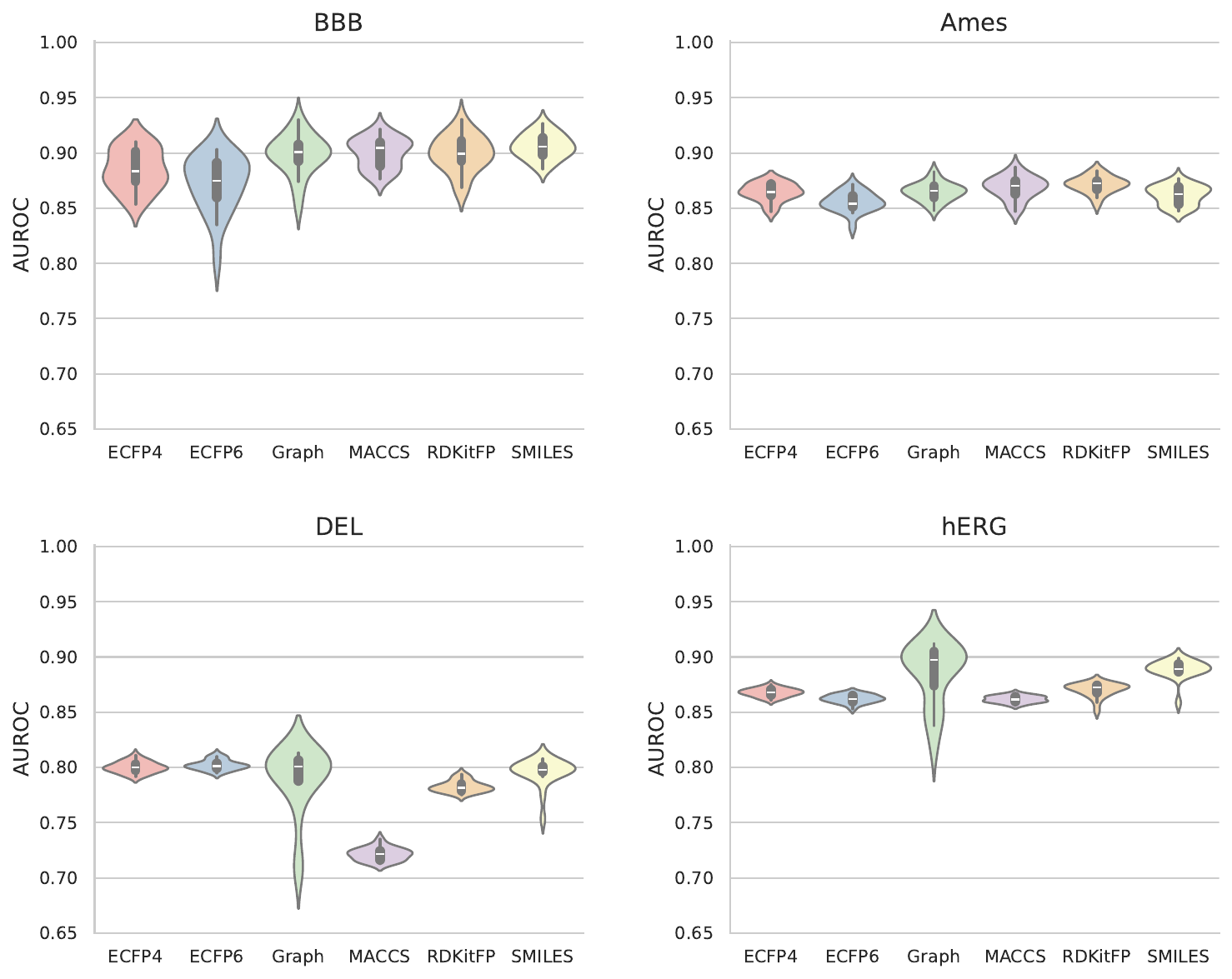} 
    \caption{\textbf{Classification performance of neural networks trained on different molecular representations in molecular property prediction tasks.} The performance is measured as the area under the receiver operating characteristic curve (AUROC). The performance is displayed as the distribution over 20 experiment repetitions.}
    \label{fig:performance}
\end{figure}

When comparing different molecular representations for neural networks, we found that models trained on graph representations showed the least information leakage across all datasets (\Cref{fig:FPR0}). The graph representation consistently had the lowest TPRs across all datasets and attacks, with a median TPR that was on average $66\% \pm 6\%$ lower than median TPRs of the other representations at an FPR of 0. In fact, for our larger datasets (DEL enrichment and hERG channel inhibition), models trained on graph representations were the only ones for which it was not possible to identify more training data molecules than by random guessing (\Cref{fig:FPR0}). We observed the same trend for an FPR of $10^{-3}$, where the graph representation consistently had the lowest TPRs (Supplementary Figure 2). We tested whether this was due to differences in model performance (\Cref{fig:performance}), but found no clear correlation between model performance and information leakage. For the small datasets, most of the models trained on different representations performed similarly. For the larger datasets, there were some outliers in model performances. In the DNA encoded library enrichment dataset, this included models trained on MACCS keys, which performed significantly worse than the other representations. In the hERG channel inhibition dataset this included models trained on graph and SMILES representations, which performed significantly better than the other representations. Our findings suggest that graph representations combined with message passing neural networks may offer the safest architecture in terms of data privacy, without sacrificing model performance.

\subsection*{Combining membership inference attacks}

\begin{figure}[h!]
    \centering
    \includegraphics[width=0.6\textwidth]{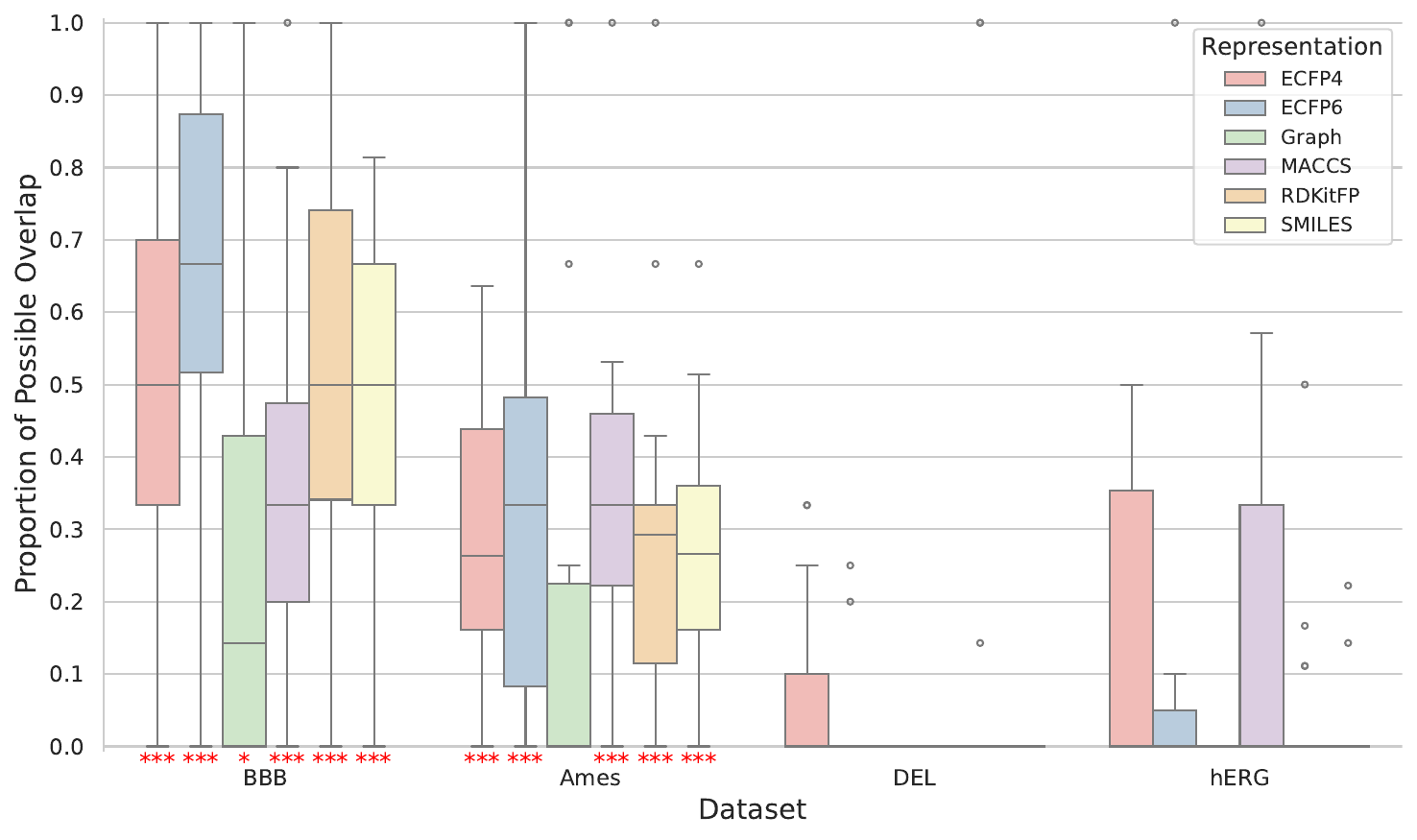} 
    \caption{\textbf{Overlap between the sets of molecules identified by the likelihood ratio attack (LiRA) and the robust membership inference attack (RMIA).} The percentage of possible overlap is defined as the proportion of molecules from the smaller set that are also present in the larger set. The less overlap exists between the attacks, the more information is gained when combining them. Overlap that was significantly higher than observed when randomly drawing two uncorrelated subsets is indicated by red stars. A single star represents a p-value less than 0.05, two stars represent a p-value less than 0.01, and three stars represent a p-value less than 0.001.}
    \label{fig:overlap}
\end{figure}

After confirming that both membership inference attacks could identify molecules from the training data, we investigated whether they identified the same molecules or whether they could be used together to gain more information about the training data. To do this, we calculated the percentage of maximum possible overlap between the sets of molecules identified by each attack (\Cref{fig:overlap}). For our small datasets, we observed significantly higher overlap than would have been observed by chance if the attacks were completely uncorrelated. However, the overlap was still well below 100\%, indicating that using both attacks can identify a wider range of molecules in the training data. For our larger datasets (DEL enrichment and hERG inhibition), there was no significant overlap, which is reasonable given our earlier findings that only one of the attacks significantly outperformed random guessing in each dataset. How much the observed overlap deviated from overlap occurring due to chance is shown in Supplementary Figure 3.  Our results suggest that using multiple different membership inference attacks is advantageous and allows the identification of more molecules from the training data.

We also investigated the overlap of identified molecules in models trained on different representations. We found a consistently large overlap between models trained on ECFP4 and ECFP6. For other representations, the overlap varied depending on the dataset and the attacks used. Detailed results can be found in Supplementary Figure 4.

\subsection*{Analysing the identified training data molecules}

Next, we wanted to see if the molecules identified from the training data shared any common characteristics. To do this, we analysed whether they differed in their distributions of property labels and molecular sizes compared to the overall training data. For the property labels, we found that the identified molecules had a significantly higher proportion of minority class molecules compared to the overall dataset (\Cref{tab:classimbalance}). This significant difference in label distribution was observed in all our imbalanced datasets and held true for both small datasets (blood-brain barrier crossing) and larger ones (DNA encoded library enrichment, hERG channel inhibition) across both membership inference attacks. We confirmed this finding by examining the TPRs of minority class molecules and discovered that their TPRs were consistently higher than the overall TPRs (Supplementary Figure 5). Specifically, the median TPR of the minority class was approximately three times greater for all representations of the blood-brain barrier crossing dataset and up to 20 times greater for some representations of the DNA encoded library enrichment and hERG channel inhibition datasets. Detailed TPR distributions for all datasets and representations can be found in Supplementary Figure 5. Regarding molecular sizes, we only found differences between identified and not identified structures in models trained on ECFP representations (Supplementary Figure 6). For models trained on other representations, we did not find any significant differences. While the identified structures do not seem to show a clear trend regarding their molecular size, our findings do indicate that it is easier to identify molecules from the minority class.

\begin{table}
    \centering
    \caption{\textbf{Property label distributions of the identified molecules and the overall datasets.} The amount of positive compounds in each dataset is written in parentheses in the 'Dataset' column. The numbers in the 'Mean' columns refer to the percentage of positive compounds in the identified molecules. Stars indicate significant differences in the property label distribution of the identified molecules compared to the property label distribution in the training data. A single star represents a p-value less than 0.05, two stars represent a p-value less than 0.01, and three stars represent a p-value less than 0.001.}
    \begin{tabular}{l>{\raggedright\arraybackslash}l
      S[table-format=1.2]
      S[table-format=1.2]
      S[table-format=1.2]
      S[table-format=1.2]}
    \toprule
    & & \multicolumn{2}{c}{LiRA} & \multicolumn{2}{c}{RMIA} \\
    \cmidrule(lr){3-4} \cmidrule(lr){5-6}
    Dataset & {Representation} & {Mean} & {Significance} & {Mean} & {Significance} \\
    \midrule
    \multirow{6}{*}{\makecell[l]{BBB\\ (0.76)}} 
        & ECFP4 & 0.16 & *** & 0.25 & *** \\
        & ECFP6 & 0.07 & *** & 0.17 & *** \\
        & Graph & 0.40 & ** & 0.42 & ** \\
        & MACCS & 0.31 & *** & 0.37 & ** \\
        & RDKitFP & 0.19 & *** & 0.31 & *** \\
        & SMILES & 0.21 & *** & 0.20 & *** \\
    \midrule
    \multirow{6}{*}{\makecell[l]{Ames\\ (0.54)}} 
        & ECFP4 & 0.54 &  & 0.45 & * \\
        & ECFP6 & 0.49 &  & 0.45 &  \\
        & Graph & 0.51 &  & 0.77 & ** \\
        & MACCS & 0.50 &  & 0.53 &  \\
        & RDKitFP & 0.60 &  & 0.47 &  \\
        & SMILES & 0.44 &  & 0.44 & * \\
    \midrule
    \multirow{6}{*}{\makecell[l]{Del\\ (0.05)}} 
        & ECFP4 & 0.16 &  & 0.78 & *** \\
        & ECFP6 & 0.12 &  & 0.82 & *** \\
        & Graph & 0.00 & *** & 0.43 &  \\
        & MACCS & 0.23 &  & 0.69 & *** \\
        & RDKitFP & 0.14 & ** & 0.62 & *** \\
        & SMILES & 0.05 & ** & 1.00 & *** \\
    \midrule
    \multirow{6}{*}{\makecell[l]{hERG\\ (0.04)}} 
        & ECFP4 & 0.80 & *** & 0.55 &  \\
        & ECFP6 & 0.44 &  & 0.47 &  \\
        & Graph & 0.29 &  & 0.53 &  \\
        & MACCS & 0.75 & *** & 0.78 & *** \\
        & RDKitFP & 0.66 & *** & 0.76 & *** \\
        & SMILES & 0.72 & *** & 1.00 & *** \\
    \bottomrule
    \end{tabular}
    \label{tab:classimbalance}
\end{table}

\subsection*{Case study}

To illustrate our results, we present a specific example of attacking one neural network model trained to predict whether molecules can pass the blood-brain barrier. Molecules are represented by ECFP4s, a common representation in many related applications. This particular model was chosen because it is representative of the 20 experimental repetitions we conducted, with its TPR falling within the interquartile range of our results. Figure 5 shows the chemical structures identified using LiRA on this model under the most stringent conditions (an FPR of 0). It was possible to identify 23 of the 859 structures from the training data (\Cref{fig:mols}). The baseline for random guessing in that case is identifying 2 of 859 structures (See Supplementary information). 21 of the 23 identified structures are from the minority class (\Cref{fig:mols}). When we relaxed the FPR to $1.1 \times 10^{-2}$
(allowing for 10 false positives among the identified structures), we were able to identify 100 structures from the training data (baseline for random guessing is 10 structures in that case). This illustrates the rapid increase in identified structures as the restrictions on the FPR are relaxed. Additionally, when combining both LiRA and RMIA, we identified 53 structures at an FPR of 0. We hope that this concrete illustration shows the potential risks that membership inference attacks pose to neural network models used in drug discovery.

\begin{figure}[h!]
    \centering
    \includegraphics[width=\textwidth]{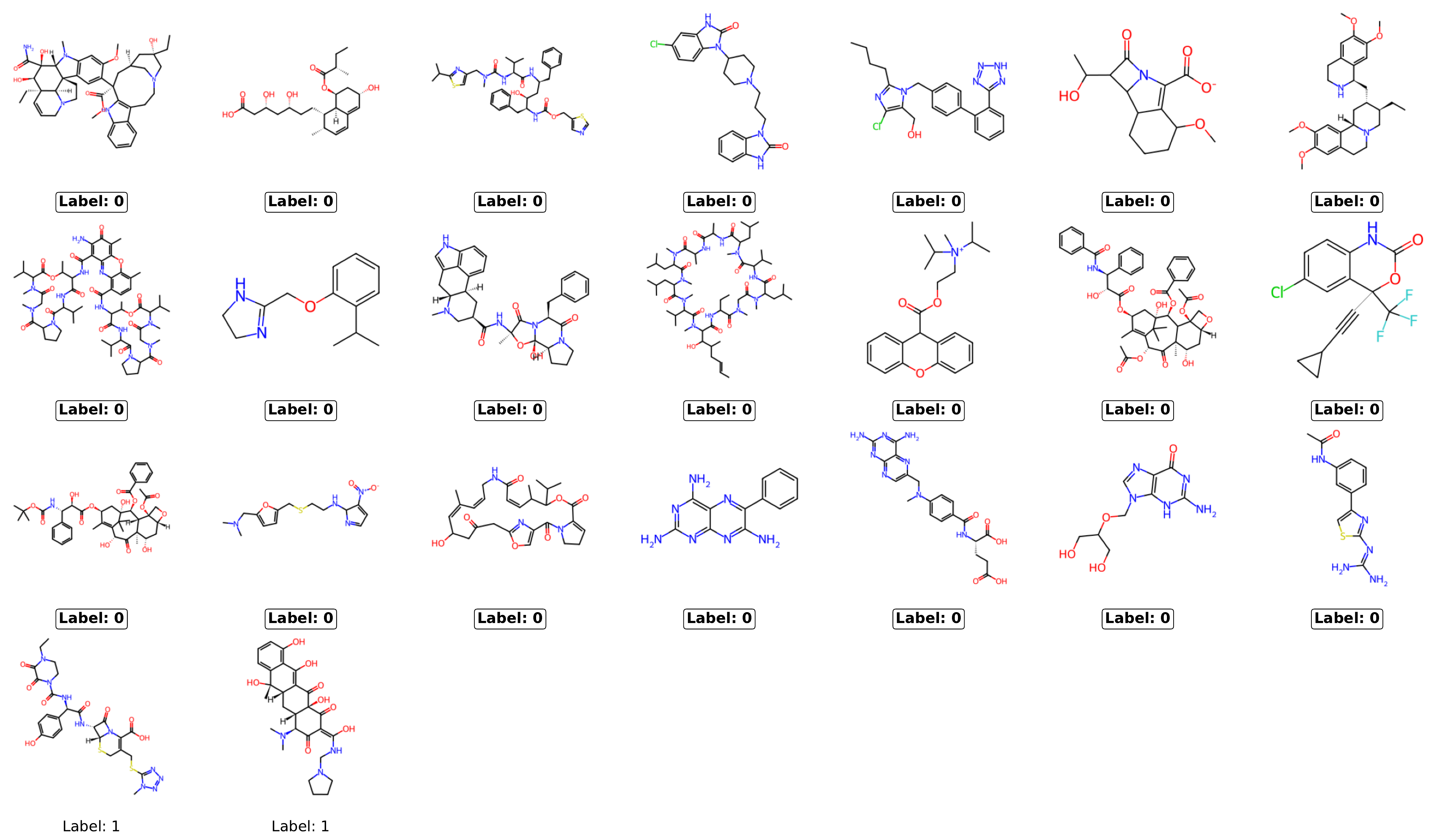} 
    \caption{\textbf{Chemical structures identified using the likelihood ratio attack (LiRA) against a neural network model trained to predict whether molecules pass the blood-brain barrier.} Molecules were represented using ECFP4s in this model. Structures that are from the minority class have the label 0 and are  surrounded by a solid line. These structures correspond to molecules that cannot pass the blood-brain barrier.  It was possible to identify 23 of the 859 training structures at an FPR of 0.}
    \label{fig:mols}
\end{figure}

\section*{Discussion}

We investigated if it is possible to identify molecules from the training data only using the output of trained neural networks, a so-called black-box attack scenario.  To investigate this question, we have applied state-of-the-art membership inference attacks to neural networks trained on different machine learning tasks for molecular property prediction. We showed that it is possible to confidently identify a subset of the training data. We also showed that combining multiple different membership inference attacks allows us to identify even more molecules since each attack identifies different molecules. Furthermore, we investigated the identified molecules and found that they contain a much higher proportion of molecules from the minority class. Thus the investigation presents  evidence that there can be significant information leakage of chemical structures from the training data when publishing a trained neural network model, which we will discuss in the following section.

It is important to note that our results focus on membership inference attacks against neural networks trained on classification tasks. This investigation does not cover regression tasks. Further research is needed to explore this area.

A limitation of the membership inference attacks we used is that they require the adversary to have data similar to the training data of the target model. In a real world scenario, this might often be the case. For many tasks in drug discovery, there are some small publicly available datasets \cite{huang2021therapeutics}. Additionally, many organisations have their own internal datasets for these tasks. Furthermore, Shokri et al. showed that even when similar data is not available, synthetic data generated from the target model can be used to successfully perform attacks that require shadow models \cite{shokri2017membership}.

We also want to emphasize that membership inference attacks assess whether it is possible to identify samples from the training data, not whether it is possible to reconstruct the training data from the model. These attacks are commonly used to assess information leakage in privacy assessments and viewed as a building block towards other attacks, e.g. reconstruction attacks \cite{salem2023sok}. In the context of drug discovery, they may have even more practical applications. For example, if an organisation offers neural network based molecular property predictions as an online service, membership inference attacks could determine whether specific molecules were part of the model's training data. Since the presence of a molecule in the training data suggests that it is being actively researched, a competitor could use this information to gain valuable insights that could give them a strategic advantage.

Our study shows that neural networks trained for molecular property prediction in drug discovery can leak training data information, as demonstrated through membership inference attacks. However, message-passing neural networks using graph representations of molecules showed significantly reduced vulnerability to these attacks. We argue that this shows that these models are the safest architecture in terms of privacy conservancy of the training data in our setting. An alternative interpretation could be that message-passing neural networks are not inherently safer, but rather that the specific membership inference attacks we used were less effective against this particular combination of model and representation. However, we think this is very unlikely, as the results are held across two different attacks, both of which rely only on model outputs rather than architecture-specific features. The only way the attacks are influenced by the specific architecture is through the training of shadow models that share the architecture of the target model. Notably, LiRA and RMIA are robust to mismatches in shadow model architectures, as shown by Carlini et al. and Zarifzadeh et al. \cite{carlini2022membership, zarifzadeh2024low}, meaning that variations in shadow model architectures do not significantly affect the success of the attacks. This supports our claim that graph representations of molecules with message-passing neural networks are the safest architecture in terms of protecting training data privacy in drug discovery.

We are confident that our results would be similar even if attacks tailored to graph classification neural networks, such as those proposed by Wu et al. \cite{wu2021adapting}, were used. Our conclusion is supported by Zarifzadeh et al. \cite{zarifzadeh2024low}, who showed both theoretically and empirically that the Attack-P method of Ye et al. \cite{ye2022enhanced} --- which is essentially identical to the threshold-based attack of Wu et al. --- is less effective than both LiRA and RMIA. Therefore, we focused on the use of RMIA and LiRA, as they are widely recognised as state-of-the-art techniques in the field and can be applied to any model architecture.

Our findings align with those of Zarifzadeh et al., who investigated membership inference attacks in the domains of computer vision (using CIFAR-10, CIFAR-100, and CINIC-10 datasets) and tabular data (using the Purchase-100 dataset) \cite{zarifzadeh2024low}. At a false positive rate (FPR) of 0, they reported true positive rates ranging from 0.0082 to 0.0778, which is in the same range as our results. This indicates that the findings of attacks on neural networks in other deep learning fields translate into the field of molecular property prediction. Another finding of Zarifzadeh et. al was that RMIA consistently outperformed LiRA \cite{zarifzadeh2024low}. They derived this both theoretically and empirically. Our results generally support this, with one exception: for attacks on the hERG channel inhibition dataset, LiRA outperformed RMIA at an FPR of 0. However, at an FPR of $10^{-3}$, this was not the case. At an FPR of $10^{-3}$, our results completely agreed with the findings of Zarifzadeh et al. \cite{zarifzadeh2024low}. The small discrepancy at an FPR of $0$ may be due to the computational constraints we faced with the hERG dataset, which was the largest in our study. Due to its large size, we had to use a small amount of samples $Z$ from the underlying distribution to do the likelihood ratio test against for RMIA. This limitation arose because comparing all data points against many points $Z$ across our models was computationally prohibitive. In contrast, LiRA does not have these constraints, which may explain its better performance compared to RMIA in this case. While RMIA generally outperforms LiRA, the latter remains a valuable approach, as it identifies different molecules, making it a complementary method, which we will discuss in a later paragraph.

Our results also show that membership inference attacks are most effective on smaller datasets. This is consistent with the findings of Shokri et al. \cite{shokri2017membership}, who link the success of attacks to the generalisability of the model and the diversity of the training data --- both of which improve with larger datasets. It is important to note that our neural networks are by no means designed in a way that makes them vulnerable to attack. On the contrary, we have implemented robust regularisation techniques that have been shown to make neural networks more resilient to membership inference attacks and improve privacy guarantees. In particular, our models use early stopping, dropout, and L2 weight regularisation. For the latter two, it has been specifically shown to reduce the efficiency of membership inference attacks \cite{shokri2017membership, jain2015drop}.

In practice, it could even be possible to increase the effectiveness of the attacks further by augmenting the attack query with some similar data as was shown by Zarifzadeh et al. \cite{zarifzadeh2024low}. We did not explore this due to computational limitations and given the broader scope of our study.

We found that by applying multiple membership inference attacks, we were able to identify more molecules within the training data. This is consistent with previous work by Ye et al. \cite{ye2022enhanced}, which demonstrated that some data points are only identified by certain attacks. We extended this by investigating the current state-of-the-art methods, LiRA and RMIA, and explicitly quantifying the overlap between these attacks across different datasets. From a practical point of view, using both attacks makes sense because it is possible to reuse the same shadow models between attacks, allowing more training data to be identified with limited computational overhead. In addition, the attacks remain feasible even when minimal computational resources are available. For example, RMIA has been shown to perform effectively with as few as two shadow models \cite{zarifzadeh2024low}. In such cases, the attacks can be run on any device capable of training neural networks with architectures similar to the target model.

Our finding that molecules in the minority class are more likely to be identified could be explained by the lower diversity in the training data for these compounds, as discussed above. This observation has important implications for drug discovery. In many datasets, the pharmacologically relevant compounds often belong to the minority class. For example, in high-throughput screening assays such as DNA-encoded library enrichment, researchers focus on the few molecules that bind to the target protein, while the majority that do not bind are of less interest \cite{satz2022dna}. This pattern is also seen in various cell-based screening assays, such as phenotypic assays aimed at identifying molecules that inhibit cancer cell proliferation \cite{zheng2013phenotypic}. In these scenarios, the minority class contains the compounds of greatest interest, making their identification far more valuable.

To address these privacy concerns, we have developed a Python package to assess the privacy of training data for molecular property prediction\footnote{\url{https://github.com/FabianKruger/molprivacy}}. This package allows users to evaluate their own data by applying our workflow to determine the extent to which training data can be identified when using different molecular representation methods. In addition, the package supports the testing of new representation methods by providing insight into their training data privacy and model performance on both user-provided and pre-supplied datasets. We hope that this tool will help researchers assess privacy risks before publishing their models.

Our research shows the potential dangers of information leakage from training data when publishing a trained neural network for drug discovery tasks. This risk exists even when the weights of the neural network are not published, and the model is offered as a supposedly safe web service. This has significant implications for organisations, which must constantly balance the need to make scientific discoveries openly available with the imperative to protect confidential data. We have shown that information leakage is consistently observed, but  it can be mitigated by representing molecules as graphs and using message-passing neural networks, which also proved to be among the best performing models on our datasets. However, when planning to publish a model, it is crucial to consider not only performance but also the privacy implications of different model architectures. Our findings also open up new research questions, such as how to adapt reconstruction attacks to the domain of molecules and how to develop models that are safer in terms of training data privacy in this field. The baseline for developing safer models might be to represent molecules as graphs and use message-passing neural networks for predictions. Our research highlights the essential balance between publicly available innovation and privacy, a balance that will impact the future of AI-driven drug discovery. 

\section*{Methods}

In this section, we first describe how we trained neural networks on biological datasets to predict molecular properties. Then, we outline the membership inference attacks used to evaluate the vulnerabilities of the models. Finally, we explain the methods used to compare and analyse the molecules leaked by these attacks. A high-level overview of our workflow is presented in \Cref{fig:workflow}. The code for our models and membership inference attacks, along with the datasets used in this study, are available on GitHub. \begin{center}
    \url{https://github.com/FabianKruger/molprivacy}
\end{center}

\subsection*{Datasets}

We used four different datasets to predict pharmacologically relevant molecular properties. The datasets differ in size, task, and class imbalance. The first dataset is used for mutagenicity prediction \cite{hansen2009benchmark, xu2012silico}. It contains Ames test results for 7,255 drugs. Of these, 54\% show positive results. The second dataset assesses blood-brain barrier permeability \cite{martins2012bayesian}. It contains 1,909 molecules, with 76\% able to penetrate the barrier. The third dataset provides information on the inhibition of the potassium ion channel encoded by the human Ether-à-go-go-Related Gene (hERG) \cite{du2011hergcentral}. Inhibition is defined as a half-maximal inhibitory concentration of less than 10 µM. This dataset contains 306,341 compounds, with 4.5\% being inhibitors. These three datasets were obtained from Therapeutics Data Commons \cite{huang2021therapeutics}. The fourth dataset contains information on whether a molecule is enriched in a DNA-encoded library (DEL) for binding to carbonic anhydrase IX \cite{lim2022machine}. Positive enrichment is defined as the top 5\% of enrichment scores. This dataset includes 108,528 molecules, with 4.9\% showing enrichment after cleaning the data.

We pre-processed all datasets to remove ambiguities and incorrect compounds. Molecules were standardised for correct bonding, aromaticity, and hybridisation. Salts were removed to isolate the primary compound and Simplified Molecular Input Line Entry System (SMILES) \cite{weininger1988smiles} strings were converted to their canonical forms. Duplicate molecules and those with conflicting labels were removed. Molecules with canonical SMILES strings longer than 200 characters were also excluded. These steps were performed using the RDKit package version 2024.03.1. The reported dataset sizes are after cleaning. The cleaned datasets were randomly divided into a training set (45\%), a validation set (10\%), and a population subset (45\%). The population subset was used for membership inference attacks, while the training and validation sets were used for model training and hyperparameter optimisation.

\subsection*{Model architectures}

To capture the variety in molecular representation approaches, we trained neural networks on a range of commonly used representations. Our study included extended-connectivity fingerprints (ECFPs) \cite{rogers2010extended}, molecular access system (MACCS) keys \cite{durant2002reoptimization}, graph representations, RDKit fingerprints (RDKitFPs) \cite{rdkit}, and SMILES \cite{weininger1988smiles} representations. We chose these representations to cover various conceptually different approaches to molecular representation. For ECFPs, we investigated fingerprints with radii of 2 and 3, both mapped to 2048-bit vectors. MACCS keys were represented as binary vectors, indicating the presence or absence of 166 structural patterns. RDKitFPs identified all subgraphs in the molecule up to a length of 7, hashed into 2048-bit vectors. These three representations were generated using RDKit \cite{rdkit}. The graph representation was generated using Chemprop version 1.6.1 \cite{yang2019analyzing}.

The type of neural network we used varied depending on the specific molecular representation. We used multi-layer perceptrons (MLPs) for ECFPs, MACCS keys, and RDKitFPs. We employed message passing neural networks implemented in Chemprop for the graph representation. For the SMILES representation, we used a pre-trained transformer encoder combined with a convolutional neural network based on Karpov et al. \cite{karpov2020transformer}. All our models were implemented in Pytorch version 2.2.2 \cite{paszke2019pytorch}. We pre-trained the transformer encoder to convert non-canonical SMILES strings to their canonical counterparts for 20 epochs using the ChEMBL\_V29 dataset from Therapeutics Data Commons \cite{zdrazil2024chembl}. We randomly split this dataset into 90\% training data and 10\% validation data. For the transformer encoder, we used the same hyperparameters as in the original publication but increased the context length of the transformer from 110 to 202 tokens in order to also generate encodings for larger molecules. We determined the hyperparameters for the MLPs, message passing neural networks, and convolutional neural networks using a Bayesian optimisation method, which we will describe in the next paragraph. All our models had one output node to predict the logits for our binary classification problems.

\subsection*{Hyperparameter optimization}

To avoid introducing subjective bias into our models, we decided to automatically optimise the hyperparameters of the neural networks using a tree structured Parzen estimator \cite{bergstra2011algorithms}. This was done using Optuna version 3.6.0. \cite{akiba2019optuna}. We optimised dropout rate, number and dimension of hidden layers, learning rate, and weight decay for MLPs. For message passing neural networks, we optimised message passing steps, dropout, encoder hidden dimension, bias addition in the encoder, aggregation function, number and dimension of classifier hidden layers, learning rate, and weight decay. For convolutional neural networks, we kept the filter sizes from the original publication and optimised dropout, learning rate, and weight decay. Detailed ranges for the hyperparameter search spaces are shown in Supplementary Table 1. We optimised each neural network architecture for three hours on an NVIDIA Volta V100 GPU. During this time, we evaluated the validation cross-entropy loss for different hyperparameter combinations. Each training run was performed for a maximum of 20 epochs. We stopped runs early if the validation loss did not improve for two consecutive epochs or if, after 15 epochs, the validation loss was below the median value for that epoch.

\subsection*{Model training}

After finding the optimised hyperparameters, we trained the final models until their performance converged on the validation set. We used early stopping with a patience of 10 epochs and saved the model weight of the epoch with the lowest validation loss. For all our models, we used a weighted binary cross-entropy loss as a loss function. The weights accounted for the class imbalance and were inversely proportional to the frequency of the classes. We used the adaptive moment estimation with decoupled weight decay regularization (AdamW) optimiser for MLPs and message passing neural networks \cite{loshchilov2017fixing}. For convolutional neural networks, we used the original adaptive moment estimation (Adam) optimiser to remain consistent with the original implementation \cite{kingma2014adam}. Training was done in batch sizes of 64 samples. We repeated our experiment 20 times for each dataset and representation to capture the marginal distribution of all randomness in the experiment, including dataset splitting, hyperparameter optimisation, and model weight initialisation. We examined the performance of each model on the population sample dataset (\Cref{fig:workflow}), as it was not used in any way for training or hyperparameter optimisation, and testing the performance of the model is independent of the membership inference attacks.

\subsection*{Membership inference attacks}

To determine whether an adversary can discriminate between molecules that are in the training data and those that are not, we applied two state-of-the-art membership inference attacks: likelihood ratio attacks (LiRA) \cite{carlini2022membership} and robust membership inference attacks (RMIA) \cite{zarifzadeh2024low}. Both methods assign a score to each sample, indicating the confidence that it was part of the training dataset. LiRA performs a likelihood ratio test by comparing the likelihood of the model output when the sample is included in the training dataset against when it is not (\Cref{alg:lira}). To approximate these likelihoods, so-called shadow models are trained on data from the same distribution. Some shadow models include the target sample in their training data, while others do not. We used random subsets containing 50\% of our population subset dataset to train 10 shadow models for each target model. Each target model training data sample was included in some shadow models and excluded from others. The shadow models had the same hyperparameters as the target model and were trained for 15 epochs. For each target sample, two Gaussian distributions of the rescaled output logits are modelled: one for shadow models that included the target sample in their training data and one for those that did not. The likelihood of observing the rescaled output logits of the target model is then calculated for each distribution. The ratio between these likelihoods represents the likelihood ratio that the target sample was in the training data. For more details on LiRA, we refer the reader to the original publication \cite{carlini2022membership}. 

\begin{algorithm}[h] 
\caption{\textbf{Likelihood Ratio Attack (LiRA)} tests whether a specific target data point $m$—in our case, a molecular structure $x$ with the corresponding label $y$—was part of the training data for a target neural network model $f_{\theta}$. In this attack, $N$ shadow models are trained on data drawn from a distribution similar to that of $f_{\theta}$'s training data (in our case, a similar chemical space). Some shadow models include $m$ in their training data, while others do not. The re-scaled confidence of each shadow model when predicting $m$ is then calculated. These confidences are modeled as two Gaussian distributions: one for the shadow models that included $m$, and one for those that did not. Finally, we determine whether the confidence of the target model $f_{\theta}$ is more likely to belong to the distribution of models that included $m$ or the distribution of models that did not. The likelihood ratio between these distributions, combined with a decision threshold $t$, determines whether $m$ is predicted to have been part of $f_{\theta}$'s training data.}
\label{alg:lira}
\begin{algorithmic}[1]
    \State \textbf{Input:} Target model $f_{\theta}$, Target data point $m = (x,y)$, Data distribution $\Pi$, Number of shadow models $N$, Decision threshold $t$
    
    \tikzmark{start1}
    \State Define re-scaling function: $\Phi(p)=\log\left( \frac{p}{1-p} \right)$
    \State $O_{in} \leftarrow \emptyset$
    \State $O_{out} \leftarrow \emptyset$
    \For{$i = 1$ to $N$}
        \State Sample dataset from $\Pi$: $D_i \sim \Pi^n$
        \State Flip a fair coin: $c_{i} \sim \{0, 1\}$
        \If{$c_{i} = 1$}
            \State Include $m$ in $D_i$
        \Else
            \State Exclude $m$ from $D_i$
        \EndIf
        \State Train shadow model $s_i$ on $D_i$ 
        \State Calculate re-scaled shadow model confidence for m: $o_{i} \leftarrow \Phi(s_{i}(x)_{y})$ 
        \If{$c_{i} = 1$}
            \State $O_{in} \leftarrow O_{in} \cup \left\{o_{i}\right\}$
        \Else
            \State $O_{out} \leftarrow O_{out} \cup \left\{o_{i}\right\}$
        \EndIf
    \EndFor
    \tikzmark{end1}

    \tikzmark{start2}
    \State Calculate $\mu_{in}, \sigma^2_{in}$ from $O_{in}$
    \State Calculate $\mu_{out}, \sigma^2_{out}$ from $O_{out}$
    \tikzmark{end2}
    
    \tikzmark{start3}
    \State Calculate re-scaled target model confidence for m: $o_{target} \leftarrow \Phi(f_{\theta}(x)_{y})$
    \State Calculate likelihoods:
    \State \hspace{1.5cm} $L_{in} \leftarrow N(o_{target} \mid \mu_{in}, \sigma^2_{in})$
    \State \hspace{1.5cm} $L_{out} \leftarrow N(o_{target} \mid \mu_{out}, \sigma^2_{out})$
    \State Calculate likelihood ratio: $LR \leftarrow \frac{L_{in}}{L_{out}}$
    \tikzmark{end3}
    
    \tikzmark{start4}
    \If{$LR > t$}
        \State Predict that data point $m$ was in $f_{\theta}$'s training data
    \Else
        \State Predict that data point $m$ was not in $f_{\theta}$'s training data
    \EndIf
    \tikzmark{end4}
\end{algorithmic}

\begin{tikzpicture}[remember picture,overlay]
    \def\xshift{-2.3cm} 

    \coordinate (algRight) at ($(current page.west) + (\textwidth,0)$);

    \coordinate (braceX) at ($(algRight) + (\xshift,0)$);

    \draw[decorate,decoration={brace,amplitude=5pt}]
        (braceX|-start1) -- (braceX|-end1)
        node[midway,right=4pt] {\textbf{Train Shadow Models}};

    \draw[decorate,decoration={brace,amplitude=5pt}]
        (braceX|-start2) -- (braceX|-end2)
        node[midway,right=4pt] {\textbf{Fit Gaussian Distributions}};

    \draw[decorate,decoration={brace,amplitude=5pt}]
        (braceX|-start3) -- (braceX|-end3)
        node[midway,right=4pt] {\textbf{Calculate Likelihood Ratio}};

    \draw[decorate,decoration={brace,amplitude=5pt}]
        (braceX|-start4) -- (braceX|-end4)
        node[midway,right=4pt] {\textbf{Predict Membership}};
\end{tikzpicture}

\end{algorithm} 

RMIA compares the likelihood ratio of observing the target model $f_{\theta}$ after applying the training algorithm $T$ with two different conditions: first, when the target sample $m$ is included in the training dataset $D$, and second, when a random, different, sample $z$ is included instead. This process is repeated with many different random samples. RMIA then attempts to calculate the probability that these likelihood ratios exceed a threshold gamma.

\begin{equation*}
    \text{Score}(m, f_{\theta}) \approx P_{z \sim \Pi}\left( \frac{P(F_{\Theta} = f_\theta|f_\theta = T(D \cup \left\{m\right\})}{P(F_{\Theta} = f_\theta|f_\theta = T(D \cup \left\{z \right\})} \geq \gamma \right) 
\end{equation*}

In our experiments, we chose a gamma value of 2. It was shown that the attack is robust to different values for gamma \cite{zarifzadeh2024low}.  Each likelihood of the ratio is calculated using Bayes’ rule (for brevity we abbreviate the conditions on both probabilities with $m$ and $z$ here). 

\begin{equation*}
    \frac{P(f_{\theta}|m)}{P(f_{\theta}|z)} = \left( \frac{P(m|f_{\theta})}{P(m)} \right)\cdot\left( \frac{P(z|f_{\theta})}{P(z)} \right)^{-1}
\end{equation*}

The probability $P(m|f_{\theta})$ is approximated by the probability of the correct class prediction and the probability $P(m)$ is approximated as the empirical mean of this over all shadow models (\Cref{alg:rmia}). The probabilities for the random points $Z$ are computed similarly. The complete implementation of this attack is shown in \Cref{alg:rmia}. For more details on RMIA, we refer readers to the original publication \cite{zarifzadeh2024low}. For this attack, we reused the shadow models from LiRA and used the 50\% of the population sample dataset not included in their training as random sample points $Z$ in the attack. We based our implementation of LiRA and RMIA on the implementation in the LeakPro repository of AI Sweden\footnote{\url{https://github.com/aidotse/LeakPro}}.

\begin{algorithm}[h] 
\caption{\textbf{Robust Membership Inference Attack (RMIA)} tests whether a specific target data point $m$—in our case, a molecular structure $x$ with the corresponding label $y$—was part of the training data for a target neural network model $f_{\theta}$. In this attack, $N$ shadow models are trained on data drawn from a distribution similar to that of $f_{\theta}$'s training data (in our case, a similar chemical space). Some shadow models include $m$ in their training data, while others do not. The probability of $m$ is approximated by averaging the correct class probability over all shadow models. Similarly, the probability of $m$ given $f_{\theta}$ is approximated as the probability of the correct class assignment by model $f_{\theta}$. The ratio between these probabilities is then calculated and compared to the ratios obtained for other points $z$. The final score is the proportion of points $z$ for which the ratio is at least $\gamma$ times higher for data point $m$. This score, combined with a decision threshold $t$, determines whether $m$ is predicted to have been part of $f_{\theta}$'s training data.}
\label{alg:rmia}
\begin{algorithmic}[1]
    \State \textbf{Input:} Target model $f_{\theta}$, Target data point $m = (x,y)$, Data distribution $\Pi$, Samples from data distribution $Z \sim \Pi^k$, Number of shadow models $N$, Likelihood ratio threshold $\gamma$, Decision threshold $t$
    
    \tikzmark{start1}
    \For{$i = 1$ to $N$}
        \State Sample dataset $D_i$ from $\Pi$ such that $D_i$ and $Z$ are disjoint: \Statex \hspace{\algorithmicindent}$D_i \sim \Pi^n(\cdot \mid D \cap Z = \emptyset)$ 
        \State Flip a fair coin: $c_{i} \sim \{0, 1\}$
        \If{$c_{i} = 1$}
            \State Include $m$ in $D_i$
        \Else
            \State Exclude $m$ from $D_i$
        \EndIf
        \State Train shadow model $s_i$ on $D_i$ 
    \EndFor
    \tikzmark{end1}    

    \tikzmark{start2}
    \State $P(m) \approx \frac{1}{N} \sum_{i}s_{i}(x)_{y}$
    \State $P(m|f_{\theta}) \approx f_{\theta}(x)_{y}$
    \State $\text{Ratio}_m \leftarrow \frac{P(m|f_{\theta})}{P(m)}$
    \tikzmark{end2}

    \tikzmark{start3}
    \State Counter $C \leftarrow 0$
    \For{each $z = (x', y') \in Z$}
        \State $P(z) \approx \frac{\sum_{i=1}^{N} \mathbb{I}[c_i = 0] \cdot s_i(x')_{y'}}{\sum_{i=1}^{N} \mathbb{I}[c_i = 0]}$
        \State $P(z|f_{\theta}) \approx f_{\theta}(x')_{y'}$
        \State $\text{Ratio}_z \leftarrow \frac{P(z|f_{\theta})}{P(z)}$
        \If{$\frac{\text{Ratio}_m}{\text{Ratio}_z} \geq \gamma$}
            \State $C \leftarrow C + 1$
        \EndIf
    \EndFor
    \tikzmark{end3}

    \tikzmark{start4}
    \State $\text{Score}(m, f_{\theta}) \leftarrow \frac{C}{|Z|}$
    \tikzmark{end4} 

    \tikzmark{start5}
    \If{$\text{Score} > t$}
        \State Predict that data point $m$ was in $f_{\theta}$'s training data
    \Else
        \State Predict that data point $m$ was not in $f_{\theta}$'s training data
    \EndIf
    \tikzmark{end5} 
    
\end{algorithmic}
\begin{tikzpicture}[remember picture,overlay]
    \def\xshift{-3.5cm} 

    \coordinate (algRight) at ($(current page.west) + (\textwidth,0)$);

    \coordinate (braceX) at ($(algRight) + (\xshift,0)$);

    \draw[decorate,decoration={brace,amplitude=5pt}]
        (braceX|-start1) -- (braceX|-end1)
        node[midway,right=4pt] {\textbf{Train Shadow Models}};

    \draw[decorate,decoration={brace,amplitude=5pt}]
        (braceX|-start2) -- (braceX|-end2)
        node[midway,right=4pt] {\textbf{Approximate Probabilities}};

    \draw[decorate,decoration={brace,amplitude=5pt}]
        (braceX|-start3) -- (braceX|-end3)
        node[midway,right=4pt] {\textbf{Calculate Individual Likelihood Ratios}};

    \draw[decorate,decoration={brace,amplitude=5pt}]
        (braceX|-start4) -- (braceX|-end4)
        node[midway,right=4pt] {\textbf{Calculate Final Score}};
        
    \draw[decorate,decoration={brace,amplitude=5pt}]
        (braceX|-start5) -- (braceX|-end5)
        node[midway,right=4pt] {\textbf{Predict Membership}};
\end{tikzpicture}
\end{algorithm}

We evaluated the success of our attacks by determining the true positive rates (TPRs) for identifying training data molecules at different false positive rates (FPRs). We focused our evaluation on low FPRs, as was recommended by Carlini et. al \cite{carlini2022membership} and is discussed in their paper in more detail. In both of our attacks, we give the adversary a training data sample with a probability of 0.67 and a non-training data sample with a probability of 0.33. The reason for this is that we did not want the training datasets for the models to become too small, while still using all the data points for the attack. This approach allowed us to use 45\% of the dataset size as training data for the target model. With a membership probability of 0.67, the baseline TPR at an FPR of 0 is $\frac{2}{N}$, where $N$ is the size of the training dataset. A detailed derivation of this baseline is provided in the Supplementary information. To determine if the attacks leak training data information, we compared the TPRs of our attacks to the baseline TPR of $\frac{2}{N}$. We tested for significance using Wilcoxon signed-rank tests over the 20 repetitions of each experiment. We repeated this experiment with the TPR at an FPR of $10^{-3}$ to see if we observe similar trends. We also investigated the ROC curves for identifying training data molecules to see the trends at all possible FPRs.

\FloatBarrier

\subsection*{Leaked molecule analysis}

We investigated whether our two membership inference attacks identify the same molecules or can be used complementarily to gain more information about the training data. To do this, we analysed the overlap between the identified molecules from each attack. In our setting, we have the training dataset $\Omega$, from which we identify two subsets, $A \subseteq \Omega$ and $B \subseteq \Omega$, each corresponding to one attack. These subsets can have different sizes and can overlap. 
We define the percentage of the maximum possible overlap as
\nopagebreak
\begin{equation*}
    f(A, B) = \frac{|A \cap B|}{\min(|A|, |B|)}.
\end{equation*}

This scalar value ranges from 0 to 1, where 1 indicates that the larger subset contains all molecules of the smaller subset. We examined the percentage of maximum possible overlap between the two membership inference attacks for every dataset and representation. For each combination, we plotted the distribution of the 20 experiment repetitions. To determine whether the overlap is significantly different from what would occur by chance when drawing two uncorrelated subsets, we calculate the difference between the expected overlap by chance and the observed overlap. This is done for our 20 experiment repetitions. We use a Wilcoxon signed-rank test to assess if the difference between the observed and random overlap is significantly different from 0. The overlap by chance can be thought of as a random variable following a hypergeometric distribution, because when we independently draw the smaller subset, we draw without replacement from the training dataset $\Omega$, which contains the larger subset as possible overlap successes. 

\begin{equation*}
    |A \cap B| \sim \text{Hypergeometric}\left(N = |\Omega|, K = \max(|A|, |B|), n = \min(|A|, |B|)\right)
\end{equation*}

The mean of the hypergeometric distribution is defined as $\mathbb{E}[|A \cap B|] = \frac{nK}{N}$, which is the overlap that is expected to be observed by chance. We also calculated the overlap between the identified molecules from models trained on different molecular representations of the same training data.

In addition, we investigated characteristics of  the molecules that could be identified. To do this, we compared the distributions of property labels of identified molecules with the underlying property label distribution. For each of the 20 experiment repetitions, we calculated the percentage of positive compounds for both identified and not identified molecules. We then used Mann-Whitney U tests to analyse whether the two distributions differed significantly. We also calculated the TPRs of the minority class. We did this similarly as before but only considered the training data molecules of the minority class in this case. We also assessed whether the identified molecules differed in size compared to the rest of the training dataset. To do this, we calculated the number of atoms in each molecule and pooled the amounts across all 20 experiment repetitions for both identified and not identified molecules. We compared the distributions of molecule sizes and determined significance using Mann-Whitney U tests.

\section*{Acknowledgement}

This study was partially funded by the Horizon Europe funding programme under the Marie Skłodowska-Curie Actions Doctoral Networks grant agreement “Explainable AI for Molecules - AiChemist”,  no. 101120466. \newline
The work of Johan Östman was funded by Vinnova, the Swedish innovation agency, under grant 2023-03000.

\section*{Conflict of interest}

The authors declare no conflict of interest.

\newpage

\bibliographystyle{unsrtnat}
\bibliography{references}  






\end{document}